\documentclass[a4paper]{article}
\usepackage{amsmath}
\usepackage{graphicx}
\begin{document}
\title{Localization of the glueball and possible decuplets of tensor 
mesons} 
\author{Micha\l \quad Majewski\thanks{e-mail: m.majewski@merlin.phys.uni.
lodz.pl}\\
Department of Theoretical Physics II, University of Lodz\\
Pomorska 149/153, 90-236 Lodz, Poland }
\maketitle

\begin{abstract} 
We continue the study of flavor aspects of strong interaction. The answer 
to the question of glueball existence is probably a necessary step in this 
investigation. The existence of the glueball implies the existence of the 
decuplet. The model we consider requires that the mass of meson decuplet 
dominated by the glueball is enclosed between the masses of the ideal 
nonet N and S states.

This description is applied to the two sets of the tensor mesons. One of 
them comprise the well known particles which are attributed to the 
widely known nonet $2^{++}$.  We argue that they belong to a decuplet 
if there exists the isoscalar tensor meson having the mass obeying the 
condition on glueball dominated meson; the meson $f_2(1430)$ could be a 
candidate. Another set includes large number of mesons lying in the 2GeV 
region. This set is used to test ability of the model to select
particles belonging to the decuplet.

\end{abstract}
\section{Introduction}
The quark-gluon picture of strong interactions and the glueball 
hypothesis \cite{Fri-Gel} posed a question of experimental confirmation 
of the glueball existence. This requires, first of all, identifying the 
isoscalar meson which is not a $q\bar{q}$ state. The most direct way to 
achieve this would be the discovery of an isoscalar meson with exotic 
signature $J^{PC}$, i.e. having such combination of J,P,C quantum 
numbers which is not allowed for the $q\bar{q}$ states. The exotic 
isoscalar meson would be the pure glueball state as it cannot mix  
with the isoscalar $q\bar{q}$. The exotic glueballs are admitted by 
some models like constituent glue or lattice QCD but were not observed 
so far. It is thus not possible to discover the glueball by studying 
the properties of a single particle. 

There exist, however, a large number of nonexotic isoscalar mesons 
attributed to various signatures $J^{PC}$.  The non-exotic glueballs 
may be hidden among them. We expect the glueball to be the one of three 
isoscalar unphysical states belonging to the same multiplet. Its 
mixing with $q\bar{q}$ isoscalar nonet states enlarges the 
multiplet of the light mesons to a decuplet. In order to extract 
the glueball state (G) from the measurable physical isoscalar states 
we must know mixing matrix (MM) of the decuplet.  The MM is a quantity 
of great importance for G search and to construct it accurate data and 
reliable procedure are required.

The $q\bar{q}$ states and the glueball are related due to the mixing. 
The pure  $q\bar{q}$ states reflect nonet properties of the mesons. 
When glueball gets mixed with the $q\bar{q}$ states, one should expect 
modification of the nonet properties. Conversely: specific properties 
of some observed nonet may suggest the influence of glueball mixing. 
Many nonets can be regarded as more or less deformed. Two of 
them:  $0^{-+} (\pi ,K ,\eta ,\eta ' )$ and $2^{++} (a_2,K^*_2,f_2,f_2')$ 
were once considered as the most promising in this respect (the former 
has bizarre mixing angle; the latter suffers with too large difference 
between $a_2$ and $f_2$ masses).

Particular interest just in these two nonets was induced by 
observations of the $\iota(1440)$ and $\Theta(1640) $ signals in the 
$J/\Psi $ radiative decays (we use original symbols). To these signals 
there were attributed the $0^{-+}$ and $2^{++}$ signatures, respectively. 
There was a hope that establishing the connection between the meson and 
the deformation of the nonet will confirm the existence of the glueball and, 
at the same time, explain the deformation mechanism \cite{Ros}. 
These attempts failed for several reasons (we comment on this later). 
The failures suggest more scrupulous examination of the procedure 
applied and improving or replacing it by a more effective one. However, 
the original idea about glueball mixing should be preserved. Thus, for 
revealing the non-exotic glueball we have to investigate ten related 
mesons which form a decuplet. Yet this should not be regarded an obstacle. 
The analysis of the mixing is the necessity in meson spectroscopy even if 
the glueballs would be already discovered.

\section{Standard diagonalization of the mass operator}
To be more specific, we examine how the unphysical isoscalar states 
$(q\bar{q})_{octet}$, $(q\bar{q})_{singlet}$ and $G$ are distributed 
among the states  of the isoscalar physical mesons $x_1$, $x_2$, $x_3$.  
The MM transforms these states from the basis where their properties are 
defined to the basis where  the  mass operator is diagonal. The way to 
determine MM is  diagonalization of the initial mass operator. We use 
quadratic mass operator; the particle symbol means its name or mass squared.

The most obvious procedure is standard diagonalization. In the simplest 
case the 2x2 matrix describing mixing of two nonet states is diagonalized. 
\begin{equation} \label{1}
\left[
\begin{array}{cc}
x_8 & \alpha\\
\alpha & x_0
\end{array}
\right]\stackrel{\text{diag}}{\longrightarrow}
\left[
\begin{array}{cc}
x_1 & 0\\
0& x_2
\end{array}
\right].
\end{equation}
Here $x _8$ is an octet isoscalar meson,  $x_0$ is a SU(3) singlet, 
$\alpha $ is a mixing parameter, $x_1$, $x_2$ are the isoscalar physical 
mesons. The mass of the octet isoscalar meson $x _8$ is assumed to be 
determined by (quadratic) Gell-Mann - Okubo (GMO) mass formula:
\begin{equation}
x _8=\frac{1}{3}a+\frac{2}{3}b;  \label{2}
\end{equation}
where $a$ is an isovector meson; $b$ (a parameter having sense of unphysical
 $s\bar{s}$ state mass squared) is given by 
\begin{equation}
    b=2K-a;    \label{3}
\end{equation}
$K$ is the strange meson.

To diagonalize the matrix we must know its elements. In the relation 
(\ref{1}) two of them -- $x_0$ and $\alpha$ are always unknown but they 
can be eliminated if the masses of $x_1$ and $x_2$ mesons are known. 
This can be done by using the invariants of diagonalizing transformation: 
$tr(m^2)$ and $det(m^2)$. For $2x2$ matrix the invariants 
are very simple:
\begin{equation}
x_8+x_0=x_1+x_2 \label{4}
\end{equation}
\begin{equation}
x_8x_0-\alpha ^2=x_1x_2 \label{5}
\end{equation}
Therefore, if the masses of physical mesons are known, no further  
information is necessary: the procedure of diagonalization is model 
independent.

The situation changes when three isoscalar states are mixed. 
We want to diagonalize 3x3 symmetric matrix

\begin{equation} \label{6}
\left[
\begin{array}{ccc}
x_8&\alpha&\beta\\
&x_0&\gamma\\
&& G
\end{array}
\right]\stackrel{\text{diag}}{\longrightarrow}
\left[
\begin{array}{ccc}
x_1&0&0\\
&x_2&0\\
&&x_3
\end{array}
\right].
\end{equation}

Using (\ref{2}) for $x_8$ and assuming that the physical mesons $x_1$, 
$x_2$, $x_3$ are known, we find that the procedure (\ref{6}) depends on 
5 unknown parameters. These parameters are related to three invariant 
functions of the diagonalizing transformation: $tr(m^2)$, $tr((m^2))^2$ 
and $det(m^2)$. Therefore, two of the unknown parameters cannot 
be eliminated. If we want to diagonalize the matrix $m^2$, we must fix 
them. For this  purpose one usually resorts to the approximations 
simplifying the mass operator. It is convenient to make the 
approximations in the ideal quark-glueball basis
\begin{equation}
N=\frac{u\bar{u}+d\bar{d}}{\sqrt{2}},\qquad S=s\bar{s},\qquad G   
\label{7}\\
\end{equation}
which helps in better understanding the physical sense of the made 
approximations. A number of such simplified operators constructed with 
various motivations (or being the response to various demands) have been 
proposed (see e.g. \cite{Fr-N});  These operators are different and 
imply different MM's. 

Unfortunately, the above approach did not appear to be reliable. There 
were many attempts to describe mixing of the $\Theta $ meson with $2^{++}$ 
nonet (see e.g. \cite{Ros}) and the $\iota$ meson with the $0^{-+}$ 
nonet \cite{Mesh}, but no definite answer has been obtained to this 
question. It became finally clear that the signature $J^{PC}=0^{++}$ 
(not $2^{++}$!) should be attributed to the  signal $\Theta $. Also it 
appears that in the neighbourhood of the $\iota $ meson there exist 
several pseudoscalar mesons constituting 
a separate multiplet. So in both these cases the mixing partners have 
been selected in a wrong way. We see that such an approach may lead to 
accidental result which cannot be treated as reliable prediction. The 
failure of these attempts have slow down the progress in studies of 
meson spectroscopy.  It is thus desirable to have a procedure which 
could help to avoid similar confusions even if only partly.   

There were also attempts to determine  the mass of a pure glueball 
(see e.g. \cite{Se-Bi}). 
The most prominent are calculations of lattice QCD \cite{Mor} predicting 
the glueball masses for different $J^{PC}$. The lowest predicted states  
are 
\begin{eqnarray} 
0^{++}(1710), \quad 0^{-+}(2560), \quad 2^{++}(2300).
\end{eqnarray}
Their masses much exceed the expected values.

\section{Master equations for light meson multiplets}  
\subsection{Conditions allowing multiplet existence}
There exists another model well describing the properties of known 
octet and nonets of $SU(3)$ broken symmetry which provides as well 
the very convenient tool for investigating the decuplets. We call it 
the model of vanishing exotic commutators (VEC) 
\footnote{formerly named the exotic commutator (ECM) model} 
(see Appendix 1.); below, the term "multiplet" means the multiplet 
of VEC-broken  $SU(3)$ symmetry. 
 
VEC is based on the postulate of vanishing of the sequence of exotic 
commutators build out of generators and their time derivatives. This 
sequence can be transformed \cite{MajT,MT} into a sequence of algebraic 
equations describing meson multiplets called the master equations (ME)  
\begin{equation}
\sum_{i} l_i^2x_i^r=\frac{1}{3}a^r +\frac{2}{3}b^r,\quad    
r=0,1,2,\ldots \label{8}
\end{equation}
where $r$ is the power index; $a$, $b$, $K$, $x_i$ have been already 
described and the 
index $i$ runs over all isoscalar mesons of the multiplet. 
The number of the isoscalar meson is growing with growing mass:   
\begin{equation}
x_i<x_{i+1}.  \label{9}
\end{equation}
The coefficients $l_i$ express octet contents of the isoscalar 
states $x_i$ 
\begin{equation}
|x_8\rangle=\sum_i l_i|x_i\rangle,    \label{10}
\end{equation}
where  $l_i$ are real numbers because the wave functions $x_8$ 
and $x_i$ describe uncharged particles. Hence,
\begin{equation}
l_i^2\geq 0 \qquad     i=1,2,\ldots  \label{11}.
\end{equation} 

This is the requirement that the octet contents $l_i^2$ of the  physical 
isoscalar states $x_i$ should  be positive.
 
The ME (\ref{8}) is a sequence of equations which are linear with 
respect to unknown variables $l_i^2$. The solution is determined by the 
masses of octet states $a$,$b$ which are indicated by experiment   
and the masses  of undefined number of physical isoscalar mesons 
$x_i$. As the $a$,$b$ are assumed to be known from experiment they are
considered as diagonal elements of the multiplet (they may be the same 
for different multiplets, e.g. nonet and decuplet); 
$x_i$ result from diagonalization of its isoscalar sector.

The functions $l_i^2$ being the solution of the ME, are not a priori 
positive; therefore we require them to satisfy the conditions 
(\ref{11}). These conditions restrict the masses of particles 
belonging to the multiplet; the particles may form multiplet only 
if  their masses satisfy these conditions.

\subsection{Number of master equations, solvability
conditions\\ and genesis of the multiplet mass formulae}
The number of equations (\ref{8}) which are to be taken into account 
is not declared in advance because it depends on the multiplet. Yet 
this number cannot be smaller than the number of variables $l_i^2$ 
since we want to describe the multiplet. So they form the minimal 
(basic) sequence of ME describing the multiplet.

The multiplet can also be described by the system of ME larger than 
the basic one. Such a system for linear equations is overdetermined. 
Hence, to be a solution the $l_i^2$'s must satisfy some extra 
solvability conditions. The $l_i^2$'s depend only on the masses of the 
multiplet particles, therefore the extra condition can be satisfied 
only at cost of the mass restriction.
 
The solvability conditions are described by the sequence (\ref{8}) as 
well. Inserting the solution of the basic sequence of ME $l_i^2$ into 
any of the subsequent equations we obtain some relation $F_j$ = 0 
which depends only on the masses of the multiplet. Hence, it constitutes 
a MF. We thus get a set of MF's  
\begin{eqnarray*}
F_j=0, \quad   j=1,2,3     \ldots\
\end{eqnarray*}  
Each of the MF's corresponds to one of the extra equations (\ref{8}) and 
puts on  the masses one restriction. As the multiplet has finite number 
of particles the number of independent relations $F_j$ = 0 complying with 
data must be finite. The sequence of equations describing the multiplet 
ends when the subsequent  $F_k$=0 contradicts the data. 

Note that the MF is not necessary for existence of the multiplet. On the 
other hand, the multiplet may have more than one MF. Since the number 
of MF is finite we can number them. If $F_1$ = 0 is satisfied by the data, 
the multiplet has at least one MF. Then we should check whether the second 
MF ($F_2$=0) is also satisfied. The multiplet has two MF's if the latter 
equation complies with the data and does not reduce to the previous 
restriction. Clearly we can prolong this procedure and check the existence 
of the $F_3$=0 and so on. 
 
However, it may also happen that starting from some $F_k=0$, the 
following MF's comply with data but reduce to the earlier result. Then 
the corresponding equations (\ref{8}) do not influent the description 
of the multiplet and may be abandoned. Hence, also in this case the 
sequence (\ref{8}) describing the multiplet is finite. 

Thus the number of ME (\ref{8}) describing the multiplet is equal to 
sum of two numbers: the number of isoscalar states and the number of 
MF's relating the masses of the multiplet. 

The multiplets having different number of MF's are described by VEC as 
different multiplets. They have the same basic sequence of ME but they 
are described by different number of equations (\ref{8}).  

We can see that the VEC model offers the description of wide variety of 
the multiplets. The study on existence the multiplets as well as the   
mechanisms of their appearance is subject of the meson spectroscopy.  
The VEC model is a promising tool for investigating these 
phenomena.

The features of the VEC description can be apparently seen on example of 
the nonets. There are three allowed kinds of the nonet which correspond 
to zero, one and two MF's obeyed by them. The procedure defining the 
kinds of nonet has been already presented \cite{MajT, enc} for other 
purpose. Now we remind it stressing the aspects which will be useful 
during investigating the decuplet.

\subsection{Description of the nonets}
The basic (minimal) sequence for the description of nonet consists of the 
first two equations (\ref{8}). Depending on the number of MF's three 
types of the nonets are distinguished which are named: the 
Gell-Mann - Okubo (GMO), Schwinger (S) and Ideal (I) for the cases
of 0, 1 and 2  MF's, respectively.

\begin{itemize}
\item  Gell-Mann -- Okubo (GMO) nonet\\ 
This type of the nonet arises as a solution of the system of the first 
two ME (\ref{8}). In this case we have no MF. The solution is
\begin{subequations}\label{12}
\begin{align}
\label{12a}
l_1^2=\frac{x_2-x_8}{x_2-x_1},\\
\label{12b}
l_2^2=\frac{x_8-x_1}{x_2-x_1};
\end{align}
\end{subequations}
or
\begin{subequations}\label{13}
\begin{align}
\label{13a}
l_1^2=\frac{1}{3}\frac{(x_2-a)+2(x_2-b)}{x_2-x_1},\\
\label{13b}
l_2^2=\frac{1}{3}\frac{(a-x_1)+2(b-x_1)}{x_2-x_1};
\end{align}
\end{subequations}
The nonet is described by the \it{mixing angle} \rm $\vartheta$ as a 
parameter. The angle $\vartheta $ is defined as
\begin{equation}
\tan^2\vartheta =\frac{l_1^2}{l_2^2}.\label{14}
\end{equation}
The solution (\ref{12}) of ME is determined by the masses. However, 
not all possible solutions describe the nonet.  The masses of the isoscalar 
mesons are constrained by the relations
\begin{equation} \label{15}
x_1< x_8< x_2
\end{equation}
which follow from the conditions (\ref{11}); here $x_8$ is GMO mass 
(\ref{2}). The GMO nonet describes pseudoscalar  mesons: $\pi $, $K$, 
$\eta $, $\eta '$. 

\item  Schwinger (S) nonet \\
This type of nonet arises by solving the system of three  ME 
\ref{8}.\\
Both the solution (\ref{13}) and the definition (\ref{14}) 
remain true, but now the masses are related by the MF  
\begin{equation} \label{16}
(a-x_1)(a-x_2)+2(b-x_1)(b-x_2)=0
\end{equation}
which is called the Schwinger MF.\\
The constraints on the masses of the S-nonet are stronger than on the 
GMO-nonet ones. It follows from (\ref{13}) and (\ref{16}) that the 
masses of S-nonet must comply with one of two possible mass ordering 
rules (MOR) \cite{Sum}:
\begin{subequations}\label{17}
\begin{align}
\label{17a}
x_1<a<x_2<b, \quad  \tan^2\vartheta <\frac{1}{2},\\
\label{17b}  
a<x_1<b<x_2, \quad  \tan^2\vartheta >\frac{1}{2}.
\end{align}
\end{subequations}
This solution describes most of the nonets independently of their 
$J^{PC}$.\\

\item   Ideal (I) nonet\\
The third type of nonet arises as the solution of the four ME (\ref{8}).\\
Apart from (\ref{16}) also the second MF emerges
\begin{equation} 
a(a-x_1)(a-x_2)+2b(b-x_1)(b-x_2)=0.  \label{18}
\end{equation}
Solving (\ref{16}) and (\ref{18}) with respect to $x_1$ and $x_2$ we find
\begin{equation} \label{19}
x_1=a,\quad x_2=b,\quad l_1^2=\frac{1}{3}, \quad l_2^2=\frac{2}{3}, \quad
\tan^2\vartheta =\frac{1}{2}.
\end{equation}
Hence, the nonet of the third type is ideal. None of the observed 
meson nonets is strictly I-nonet but many of them deviate from it only 
slightly. Perhaps the most ideal are nonets $1^{--}$ and $3^{--}$.
\end{itemize}
Further increasing of the number of equations (\ref{8}) does not change 
the solution (\ref{19}). The S-nonet has one MF; the I-nonet has two MF's. 
Their mixing angles are different.   

Notice that the equations (\ref{16}) and (\ref{18}) have also the 
solution 
 $x_1=b$ and $x_2=a$  \cite{MajT}. 
This solution does not describe any known multiplet. 
For choosing  the physical solution describing the I-nonet we need to 
know which of the $x_i$ mesons is N (or S) state (\ref{7}). The old 
choice of the solution (\ref{19}) was just the result of common consent 
that the heavier of isoscalar mesons is the hidden strange $S=s\bar{s}$ 
state. 

The same concerns the Schwinger nonet but this time we must know which 
of the observed $x_i$ mesons is \it{dominated} \rm (not completely  
determined) by N (or S) state.

The N/S domination patterns of the isoscalar states is identical for all 
nonets independently of their signature $J^{PC}$ and the number of MF's 
satisfied by them. 

The notion of \it{dominance pattern} \rm will play important role in our 
investigation of the decuplet. 

\section{Description of decuplets}
\subsection{Diagonalization and wave functions} 
The basic sequence describing the decuplet consists of three 
equations  (\ref{8}). There are two possible kinds of the decuplet: 
the one having no MF and the one having one MF.  The functions $l_i^2$ 
solving the basic sequence of ME (\ref{8}) are explicitly expressed in 
terms of the masses 

\begin{subequations}\label{20}
\begin{align}
\label{20a}
    l_1^2=\frac{1}{3}\frac{(x_2-a)(x_3-a)+2(x_2-b)(x_3-b)}
    {(x_1-x_2)(x_1-x_3)}, \\
\label{20b}
    l_2^2=\frac{1}{3}\frac{(x_1-a)(x_3-a)+2(x_1-b)(x_3-b)}
    {(x_2-x_1)(x_2-x_3)}, \\
\label{21c}
    l_3^2=\frac{1}{3}\frac{(x_1-a)(x_2-a)+2(x_1-b)(x_2-b)}
    {(x_3-x_1)(x_3-x_2)}.
\end{align}
\end{subequations}
 
From the two kinds of the decuplets the more reliable and easier to 
complete is the one having MF. We restrict our attention to it.

The MF is \cite{enc,Where}
\begin{equation} \label{21}
(x_1-a)(x_2-a)(x_3-a)+2(x_1-b)(x_2-b)(x_3-b)=0.
\end{equation}
The conditions (\ref{11}) and MF (\ref{21}) impose a number of 
constraints on the masses. These constraints  can be presented in the 
form of the mass ordering rule (MOR) (see Appendix 2.)
\begin{equation} \label {22}
x_1<a<x_2<b<x_3.
\end{equation}
This rule is an indispensable tool for selecting the candidates for 
decuplet. 

Let us introduce the mixing matrix U transforming isoscalar states of
exact symmetry $SU(3)$ into the physical ones:
\begin{equation}\label{23}
 \begin{bmatrix}
   x_1\\
   x_2\\
   x_3\\
 \end{bmatrix}   =U\begin{bmatrix}
   x_8\\
   x_0\\
    G \\
 \end{bmatrix};
\end{equation}
the initial and final states have been described in the section 2.
This matrix is orthogonal and can be written in the form

\begin{equation}\label{24}
    U=\begin{bmatrix}
     c_1 & -s_1c_2 & s_1s_2 \\
      s_1c_3 & c_1c_2c_3-s_2s_3 & -c_1s_2c_3-c_2s_3\\
      s_1s_3 & c_1c_2s_3+s_2c_3 & -c_1s_2s_3+c_2c_3\\
    \end{bmatrix},
\end{equation}
where $c_j$= cos $\vartheta_j$,  $s_j$ = sin $\vartheta_j$, ($j=1,2,3$) 
and  $\vartheta_j$ are Euler angles:\\
$0\leq\vartheta_1<\pi$; \hspace{0.2cm} $0\leq(\vartheta_2$,\quad 
$\vartheta_3) <2\pi.$

The elements of the first column are just the coefficients  $l_1$, 
$l_2$, $l_3$ introduced in (\ref{10}). Their squares are the solution 
(\ref{20}) of the ME. Therefore, we have:
\begin{equation}   \label{25}
    c_1=l_1; \;\; s_1c_3=l_2; \;\;
    s_1s_3=l_3.  
\end{equation}
Hence, the absolute values of the trigonometric functions of the Euler 
angles $\vartheta_1$ and $\vartheta_3$ are expressed in terms of masses.

To compare the predictions with data we express the MM in the basis 
of the N, S, G states. In this basis the physical isoscalar states are 
described by the matrix V
\begin{equation}\label{26}
 \begin{bmatrix}
   x_1 \\
   x_2 \\
   x_3 \\
 \end{bmatrix}=V\begin{bmatrix}
     N \\
     S \\
     G \\
 \end{bmatrix},
\end{equation}
where
\begin{equation}\label{27}
    V=UQ,
\end{equation}
and the matrix Q
\begin{equation}\label{28}
    Q=\begin{bmatrix}
      \frac{1}{\sqrt{3}} & -\sqrt{\frac{2}{3}} & 0 \\
      \sqrt{\frac{2}{3}} & \frac{1}{\sqrt{3}}& 0 \\
      0 & 0 & 1 \\
    \end{bmatrix},
\end{equation}
transforms the bases
\begin{equation}\label{29}
    \begin{bmatrix}
      x_8 \\
      x_0\ \\
      G  \\
     
    \end{bmatrix}=Q\begin{bmatrix}
      N \\
      S \\
      G \\
    \end{bmatrix}.
\end{equation}

The relations (\ref{20}), (\ref{24}) and  (\ref{25}) show that the 
matrix V depends on trhe masses of decuplet particles. However, the angle 
$\vartheta _2$ remains unknown. It cannot be determined on the basis of 
the solution of ME and is a free parameter of the MM. This freedom can be 
used for giving V some desirable specific feature. In the case of G 
mixing we require its flavor 
independence:   
 \begin{equation}\label{30}
<G|m^2|N>=\sqrt{2}<G|m^2|S>
\end{equation}
which implies
\begin{equation}\label{31}
    \tan\vartheta_2=
    \frac{c_3s_3}{c_1}\frac{x_3-x_2}{(x_3-x_1)-(x_3-x_2)c_3^2}.
\end{equation}
With this value of $\vartheta _2$ the MM depends only on the masses. 
We call it the 
\it {glueball shaped} \rm MM and label as $V_G$. Its explicit form is:

\begin{multline}
   V_G=\\
\begin{bmatrix}
      \frac{1}{\sqrt{3}} c_1-\sqrt{\frac{2}{3}} s_1c_2 & -\sqrt{\frac{2}{3}}c_1- \frac{1}{\sqrt{3}}s_1c_2& s_1s_2 \\
     \frac{1}{\sqrt{3}}s_1c_3+\sqrt{\frac{2}{3}}(c_1c_2c_3-s_2s_3) & -\sqrt{\frac{2}{3}}s_1c_3+ \frac{1}{\sqrt{3}}(c_1c_2c_3-s_2s_3)& -c_1s_2c_3-c_2s_3\\
      \frac{1}{\sqrt{3}}s_1s_3+\sqrt{\frac{2}{3}}(c_1c_2s_3+s_2c_3) & -\sqrt{\frac{2}{3}}s_1s_3+\frac{1}{\sqrt{3}}(c_1c_2s_3+s_2c_3)& -c_1s_2s_3+c_2c_3\\
   \end{bmatrix}\label{32}
\end{multline}

To have  this matrix completely determined we must fix all signs of the 
trigonometric functions $c_i$, $s_i$ (i=1,2,3) \cite{enc}. For some of 
them the sign is arbitrary due to the sign ambiguity of the matrix V but 
not for all. The other signs can be determined if N-,S-,G- 
domination is known for two of the three $x_i$ states . It 
turns out that the information about the domination has much stronger 
impact on the shape of the MM than the particular numerical values of the 
functions $c_i$, $s_i$. The former determine the character of the MM 
while the latter only change the numerical values of its elements. 
Therefore, it is important to examine the possible types of dominations 
and find out their implications and origin. 

\subsection {G-, N-, S- domination pattern of the decuplet\\
and the localization of glueball}
The genuine (not degenerated) states of the decuplet $x_i$ mesons are 
superpositions of three unphysical isoscalar states: $x_8$, $x_0$, G.  
But it is more convenient to express the states $x_i$ as linear 
combinations of N, S, G ones. The combinations must be different 
because the $x_i$ particles are different. Yet to be noticeable different, 
each isoscalar meson must be clearly dominated by one of the basis states. 
We need the information which of the isoscalar mesons $x_i$ is dominated 
by G. This information is not always available directly from experiment
\footnote{The physical meson $x_i$ dominated by G is usually called a 
"glueball candidate".\\ Observe that G in VEC model is not necessarily the 
glueball; it may be any scalar of the flavor SU(3) which is built of other 
constituents. Identification of the constituents is a separate problem.}. 
If this is the case we can try to make use of the 
properties of the remaining $x_i$ mesons (the normalization of their 
amplitudes) to infer which meson is dominated by G. We now explain how 
this can be done; we start by reminding the features of the decuplets 
containing G-state which were indicated by data.

\subsubsection {Decuplets including apparent G-dominated state}
So far two decuplet candidates including experimentally recognized 
G-dominated state have been introduced and discussed \cite{enc, 
Where}. They include the $0^{++}$ and $0^{-+}$ mesons having 
masses in the interval between 1.2GeV and 2.4GeV, which is usually 
considered as room for excited $q\bar{q}$ states. In some cases their 
existence is not firmly established and many masses are burdened by 
large errors. In spite of that, gathering these particles into a decuplet 
is possible due to MOR restrictions on the decuplet masses. \\
$\bullet$
 The  $0^{++}$  decuplet is composed of the mesons \cite{enc}:
\begin{equation} \label{33}
a_0(1450),\quad K_0(1950),\quad f_0(1370),\quad f_0(1500),\quad 
f_0(2200)/f_0(2330). 
\end{equation}
Arranging this decuplet has started after it was established that the 
meson $f_0(1500)$ is dominated by G state \cite{RPP(2010), AGST}. Three 
of the mesons (\ref{33}): $a_0(1450)$, $f_0(1370)$ and $f_0(1500)$ 
belong to "firmly established" particles \cite{RPP(2010)}. 
One of them is just the $f_0(1500)$. This supports the conclusion 
concerning the existence of the decuplet because, according to 
G-$(q\bar{q})$ mixing picture, it can be taken for granted once 
the existence of non-exotic G is established.

The enormously large difference between the masses of $K_0(1950)$ and 
$a_0(1450)$ mesons may be embarrassing but there are no other 
candidates. Moreover, the difference between the corresponding states of 
the lower lying nonet $K_0(1430)$ and $a_0(980)$ is also large. 
Therefore, we accept this difference as an experimental fact. Hence, 
we also accept their very large mass spread (difference between the 
biggest and the smallest mass of the multiplet). The MF reveals strong 
correlation between the masses 
of $K_0(1950)$ and  $x_3$ mesons. The former is measured with 
large error implying broad dispersion of the predicted $x_3$ mass. 
This is indicated in (\ref{33}) by double name of the meson 
$f_0(2200)/f_0(2330)$. Still larger uncertainty is introduced to the 
decuplet states by wide experimental dispersion of the $f_0(1370)$ mass 
which, in turn, is correlated with the mass of $a_0(1450)$ meson. The 
present data are not sufficiently accurate to make definite fit but the 
MOR restrictions considerably reduce the search range.\\ 
$\bullet$
  The decuplet $0^{-+}$ includes the mesons \cite{Where}:
\begin{equation}  \label{34} 
\pi (1300), \quad K(1460), \quad \eta (1295), \quad \eta (1405), 
\quad \eta (1475).
\end{equation} 
Its spread is small. According to  \cite{RPP(2010), AM}  the meson 
$\eta (1405)$ is dominated by G. The masses of the isoscalar mesons 
$x_i$ are measured very precisely, but the mass of the  K(1460) is 
unknown and the error for the $\pi (1300)$ is so big that it also must 
be assumed  unknown. We have at our disposal only one MF to determine 
these two masses. Moreover, the MF is a cubic equation with respect 
to each of the masses. Hence, the comprehensible solution can only be 
approximate. It was found due to MOR restrictions (\ref{22}). 
The solving procedure divides the two-dimensional domain of unknowns 
$a$ and $b$ into four disconnected sub-domains having distinct 
properties. The solution is looked for in each sub-domain separately. 
In any domain the solution of the ME is dominated by one of the $N$, 
$S$, $G$, $x_8$ states. The type of domination is preserved across each 
sub-domain. As a solution of the ME we accept the one at the sub-domain 
which assigns the G-domination to the $x_2$ state. The masses of the 
$\pi(1300) $ and $K(1460)$ mesons belonging to the sub-domain are not 
precisely determined but the ranges of their variability are of the size 
of typical experimental error \cite{Where}.

\subsubsection{Any apparent domination and localization of the glueball}
Dominations of the $f_0(1500)$ and $\eta (1405)$ mesons by G were 
established by experimental observations but not predicted by certain 
a priori postulated model. The VEC description only reveals that they 
can be understood as the components of decuplets. The common feature of 
these two decuplets is that the G dominated isoscalar meson occupies the 
same central position (of $x_2$) in the sequence of  (of $x_i$) states. 
We shall now examine implications of this statement for other decuplets. 

The G domination of the $x_2$ meson implies N, S domination of the 
$x_1$,$x_3$ mesons, respectively. The G-shaped 
MM (32) predicts 
\begin{equation}\label{35}      
|x_1> \sim |N>,  \quad  |x_3> \sim |S>. 
\end{equation}
This is standard domination pattern of the nonet a, K, $x_1$, $x_3$.

Both nonets and decuplets of mesons are described by the ME (8). The MF 
of the ideal nonet (\ref{16}),(\ref{18}) and the MF of the decuplet 
(\ref{21}) are required by consistency condition of  the overdetermined 
system of the first four equations (\ref{8}). The octet contents $l_i^2$ 
(i=1,2 - for the nonet and i=1,2,3 - for the decuplet) play role of 
unknown variables of the system; a,b are considered as known constants. 

For the I-nonet we have
\begin{equation}\label{36}  
x_1=a, \quad x_2=b, \quad |x_1>=|N>,  \quad  |x_2>= |S>. 
\end{equation}

The masses of the decuplet mesons should satisfy MF (\ref{21}). The 
structures of its isoscalar components $|x_i>$ are not ideal. However, 
if the interaction between G and isoscalar nonet states were switched 
off (by putting $l_2^2 = 0$), the ME would describe ideally mixed nonet 
with  $x_1$, $x_3$ as isoscalar components
\begin{equation}\label{37}  
x_1^{id}=a,\quad x_3^{id}=b,\quad |x_1^{id}>=|N>,\quad |x_3^{id}>=|S> 
\end{equation}
and disconnected SU(3)-singlet G.
Thus the G-shaped decuplet arises from mixing of the I-nonet with G.

The decuplet \it{mass gaps} - the bottom: $x_1^{id}-x_1$ and the upper:  
$x_3-x_3^{id}$\rm arise from mixing. They are known since the decuplet 
MF is solved. Besides, the gaps depend on spread of the decuplet. 
Hence, they are different in different decuplets but both  for $0^{++}$ 
(where the spread is large) and $0^{-+}$ (where it is small) they are 
much smaller than the I-nonet isoscalar mass difference (\ref{37})
\begin{equation} \label{38}
  x_1^{id}-x_1, \quad  x_3 -x_3^{id} \quad << \quad(x_3^{id}-x_1^{id}).  
\end{equation}
This discloses relative weakness of the G-$(q\bar{q})$ mixing mechanism 
and explains why converting the nonet into decuplet does not change the 
original N, S domination assignment of the  $x_1$,$x_3$ states. 

Obviously, the most interesting situation arises when the G-dominated 
meson among $x_i$ is not identified. Then we must first verify whether 
we are dealing with a decuplet. This can be done by checking MOR 
restrictions (\ref{22}). If it is a decuplet complying with (\ref{38}) 
then  dominations are noticeably. We thus expect for all G-shaped 
decuplets the same domination pattern  
\begin{equation}\label{39} 
  |x_1>\sim |N>,  \quad  |x_2>\sim |G>, \quad  |x_3>\sim |S>.
\end{equation}
The mass of the G dominated state $x_2$ is restricted by conditions 
\begin{equation} \label{40}
a <  x_2 <  b
\end{equation}
required by MOR (\ref{22}) and can be found by solving the decuplet MF 
(\ref{21}).

\section{Possible decuplets of $2^{++}$ mesons}
Our main purpose is to examine the existence of $2^{++}$ meson decuplets. 
The tensor mesons are very curious objects for such investigation 
because they are numerous with great majority of the isoscalar mesons.
Some of them may belong to decuplets. It is a puzzle which kind of the 
multiplets might be composed of the to date observed signals. Many of the 
recorded signals need confirmation and most of the listed masses are 
measured with low accuracy. This concerns first of all the signals 
descending from the particles having masses close to 2GeV or above. 
Therefore, predictions for the higher lying multiplet are tentative 
and should not be treated too literally. Yet, in spite of that the model 
brings transparent outline which helps to exploit the scanty, unconfirmed 
and inaccurate data on individual particles for constructing the possible 
multiplet.

The earlier studies of the $0^{-+}$ and $0^{++}$ multiplets 
\cite{enc, Where} seemed to suggest that between G and $q\bar{q}$-excited 
states there may exist a correlation resembling the chemical affinity 
\cite{Where}. It would be interesting to see how such a property could 
manifest itself in the multiplets of the $2^{++}$ mesons.

We begin with the discussion of the best known particles which are \
usually considered as nonet components.


\subsection{Is the well known tensor nonet really a nonet?} 
The currently known nonet of tensor mesons 
\begin{equation} \label{41} 
a_2(1320),\quad K_2^*(1430),\quad f_2(1270),\quad f_2'(1525)
\end{equation}       
belongs to one of the most early recognized SU(3) multiplets 
\cite{tensnonref}. The belief in its nonet status has been lasting many 
decades (excluding probably a short time of fascination with 
$\Theta (1640)$ meson). Equally stable 
- but without any doubt - is the belief that the mesons $f_2(1270)$ 
and $f_2'(1525)$ are almost pure N and S states. Perhaps such a stability 
of opinion concerning this multiplet reflects the fact that subsequent 
measurements of their masses exhibited only small changes creating no 
stimulus for reanalysis.

The present analysis was motivated by an attempt to select candidates to 
higher lying decuplet $2^{++}$, similar to the $0^{++}$ and 
$0^{-+}$ ones. This gave rise to examining all signals of tensor mesons. 
Unexpectedly, among the current data we find the ones which may change 
multiplet status of the mesons (\ref{41}). 

Two observations are especially important for reanalysis of the multiplet 
assignment of the tensor mesons (\ref{41}):   
\begin{itemize}
\item  The masses of the mesons $K_2^{*\pm }$ and $K_2^{*o}$ are known 
and individually determined. The measurement is very precise and 
difference between them much exceeds the experimental errors 
\cite{RPP(2010)} . 
\item   In few experiments within the region of appearance of the mesons 
(\ref{41}) the signals of a mysterious meson $f_2(1430)$ were recorded. 
\end{itemize}
These observations are not quite new but were abandoned previously 
because they are silent as long as the mesons (\ref{41}) are 
investigated not otherwise than as belonging to a nonet.

The present data on masses and confidence of the mesons (\ref{41}) and 
$f_2(1430)$ (a likely candidate to the decuplet) are quoted in the tab.1.
\begin{table} 
 \caption{Masses of tensor mesons belonging to a possible ground state 
decuplet. 
  Masses (in MeV) and confidence are quoted after \cite{RPP(2010)}}
  
  \vspace{0.5ex}
    \smallskip
\begin{tabular}{|c|c|c|c|c|c|}
 \hline
$\bullet a_2(1320)$ &$\bullet K_2^{*\pm}(1430)$&$\bullet K_2^{*0}(1430)$& $\bullet f_2(1270)$&
$\bullet f_2'(1525)$& $f_2(1430))$  \\
\hline
  $1318.3_{-.6}^{+.5} $ &$1425.6\pm1.5$  & $1432.4\pm1.3$ & $1275.1\pm1.2 $& $1525\pm5$&$\sim 1430$ \\
 \hline 
 \end{tabular}
 
\end{table}
In other multiplets the difference between masses of charged and neutral 
$K$-mesons are not measurable or are meaningless but in the case 
of $K_2^*$ the difference is significant and cannot be neglected. This  
raises the question which is unusual for the meson spectroscopy: which of 
the $K_2^*$ meson masses describes the breaking of SU(3) symmetry.

The dilemma is easy to solve. In the multiplet under consideration we are 
dealing with two different levels of symmetry breaking: the SU(3) and the 
SU(2) ones. The former breaking does not split the masses within $SU(2)$ 
multiplets while the latter does. The $K_2^{*0}$ is known to be modified 
by electromagnetic interaction. Therefore, the $K_2^{*\pm}$ should be 
recognized as better reflecting the properties of SU(3) broken multiplet. 
We thus define 
\begin{equation} \label{42}
K_2^* \doteq K_2^{*\pm }.
\end{equation}
However, introducing a new definition for $K_2^*$ meson mass we are 
obliged to verify its agreement with the nonet criterion (\ref{17a}). The 
MF, MOR and MM relations (which are defined for the nonet and decuplet 
masses) depend on the mass of K meson via the b parameter (\ref{3}). 
In both the nonet and decuplet multiplets the meson $f_2(1270)$ is the 
N-dominated $x_N$ isoscalar state and we can apparently see that its 
mass squared satisfies the inequality $x_N<a$. Then, keeping in mind MOR 
for  the nonet (\ref{17a}) and decuplet (\ref{22}) and assuming the 
$f'_2(1525)$ to be the S-dominated $x_S$ state we have:\\   
if $x_S<b$ we are dealing with the nonet,\\
if $x_S>b$ we are dealing with the decuplet.\\
Using the data from tab.1 and definition (\ref{3}) one can determine b. 
We should compare this value of b with the mass squared of 
the $f_2'$ meson. For such a purpose we may neglect the errors of 
$K_2^{*\pm}(1430)$ and $a_2(1320)$ in calculating b as they are much 
smaller than the error of $f_2'$ mass. We find
\begin{equation} \label{43}
b=2.3268GeV^2.
\end{equation}
The nonet MOR (\ref{17a}) requires 
\begin{equation} \label{44}
f_2' < b,
\end{equation}
while the decuplet MOR (\ref{22}) imposes 
\begin{equation} \label{45} 
b < f_2'.
\end{equation}
The mass of $f_2'$ meson regarding experimental error is
\begin{equation} \label{46} 
f_2'=(2.326\pm.015)GeV^2.
\end{equation}
Hence, within the error ranges of $f_2'$ meson mass both the inequalities 
for the nonet (\ref{44}) and for the decuplet (\ref{45}) can be obeyed. 
The question posed in the title of this subsection is thus motivated 
but no answer is obtained. We must look for the answer in 
another way.   

If the multiplet is a decuplet there should exist an extra meson which, 
together with the mesons (\ref{41}), satisfies decuplet MOR restrictions 
(\ref{22}). The decuplet MF (\ref{21}) can be solved with respect to 
the mass of this extra meson. According to the nomenclature adopted for 
the isoscalar mesons (\ref{9}) we denote this meson as $x_2$ 
(the symbols $x_1$ and  $x_3$ are attributed to the mesons $f_2(1270)$ 
and $f_2'(1525)$ dominated by N and S states). The solution is
\begin{equation} \label{47}
        x_2=m_2^2=(1,327GeV)^2 
\end{equation}
and the mesons (\ref{41}) are the components of decuplet if such a meson 
exists. 

This solution is very sensitive to some of the input masses, 
especially to  the mass of the mesons $K^{*\pm}_2$ and $f'_2$ . 
The mass (\ref{47}) is calculated on the basis of mean experimental 
values cited in the table 1. Only the mass of the  meson $f_2'$ is 
slightly changed: it is put 1526MeV instead of 1525MeV.

The MM of this decuplet is
\begin{equation}\label{48}
    V_G=\left[
          \begin{array}{ccc}
            0.8123011 & 0.3290934 & 0.481523 \\
            0.5692579 & -0.267674 & -0.777365 \\
            -0.126934 & 0.90556510 & 0.404771 \\
          \end{array}
        \right];
\end{equation}

The mass of the pure G state is 
\begin{equation}  \label{49}
m_G=1350 MeV.
\end{equation}
The signals with similar masses have been observed in several 
experiments during few decades. The bump is known as $f_2(1430)$ meson 
\cite{RPP(2010)}. The existence of this meson would support decuplet 
status of the mesons shown in the tab.1 and define it as G-dominated. 

Note that this meson was looked for in early 80-th during exploring the 
misshapen nonet of $2^{++}$ mesons \cite{Ros, R-T}. The failure of these 
attempts arise partly due to the wrongly chosen search area; the 
G-dominated meson was looked for above the mass of the meson $f'_2(1525)$ 
instead below it. The mistake was a result of scanty data: at that time 
only one measurement has been recorded below this mass. There are now further 
measurements at our disposal but the existence of the particle still is 
not "firmly established" and needs confirmation. However, now we have 
more motivation for careful reanalysis.
 
Another question may also arise. If the G-dominated meson $f_2(1430)$ 
exists one may ask why? Room for it emerges due to anomalous 
electromagnetic split of the $K^*_2(1430)$ width. Is it an accident or 
testify some correlation? No trace of correlation between 
electromagnetic and strong interactions was seen so far.

\subsection{Quest for the second tensor multiplet}
Above the ground state $2^{++}$ meson multiplet there exist 
further well established mesons and many signals waiting for 
confirmation. All these particles and signals are listed 
in the tab.2. 
\begin{table}
 \caption{Masses of tensor mesons observed above the ground state 
decuplet.\newline Masses (in MeV) and confidence are quoted after 
\cite{RPP(2010)}} 
  
  \vspace{0.5ex}
    \smallskip
\begin{tabular}{|c|c|c|c|c|c|}
 \hline
$f_2(1565)$ &$f_2(1640)$&$a_2(1700)$& $f_2(1810)$&
$f_2(1910)$& $\bullet f_2(1950)$\\ 
\hline
  $1562\pm 13 $ &$1639\pm 6 $ & $1732\pm 16$ & $1815\pm12 $& $1903\pm9$&$1944\pm 12$ \\
 \hline
 \end{tabular}
 
 {\ }\\
 
 \begin{tabular}{|c|c|c|c|c|c|}
 \hline
%
$K_2^*(1980)$ &$\bullet f_2(2010)$&$f_2(2150)$& $ f_J(2220)$&
$\bullet f_2(2300)$& $\bullet f_2(2340))$  \\
\hline
  $1973\pm 33 $ &$2011^{+62}_{-76}$& $2157\pm12$ & $2231.1\pm3.5 $& $2297\pm28$&$2339\pm 55$ \\
 \hline 
 \end{tabular}
 
\end{table}

In the tab.2 we find only one isovector $a_2(1700)$ and one isospinor 
$K_2^*(1980)$ meson. This does not fit to ten isoscalar mesons seen in 
this energy region. But the latter is poorly investigated and the 
situation may change in future. The content of the multiplet cannot be 
precisely determined, hence we focus our attention on selecting the 
particles to possible decuplet.

Only four out of the twelve signals listed in the tab.2, are "firmly 
established". Unfortunately, neither $a_2(1700)$ nor $K_2(1980)$ 
belong to this class. Yet we need definite values of their  masses to 
put MOR delimitations of the decuplet. Therefore, we accept their 
identity numbers as the values of their masses:
\begin{equation} \label{50}
 a=(1.700 GeV)^2, \quad   b=2K-a=(2.225 GeV)^2.\\
 \end{equation}
The "firmly established" mesons constitute two pairs having so close 
mass values that their difference is smaller than the error. 
Such situation is exceptional in meson spectroscopy - usually the 
difference is bigger. Using the masses from tab.2 we  see that any 
mass of the pair of the "firmly established" mesons
\begin{equation} \label{51}
f_2(1950),  \quad f_2(2010)\\
\end{equation} 
satisfy the MOR restrictions on $x_2$ 
and any particle of the pair of  "firmly established" mesons 
\begin{equation} \label{52}
 f_2(2300),  \quad f_2(2340) \\
\end{equation}
satisfies the restriction on $x_3$. 
Besides, there are also two signals  
\begin{equation} \label{53}
 f_2(1565),  \quad  f_2(1640) \\
\end{equation}
satisfying the restriction on $x_1$.

Broad limits on variables $x_i$ and the freedom of the choice of 
the $a_2(1700)$ and $K_2^*(1980)$ masses would help to fit the MOR 
restrictions as well as the MF (\ref{21}). Therefore, it is possible 
that the decuplet of these mesons exists. The mass of the corresponding 
G-dominated $2^{++}$ meson is expected to be close to 2000MeV. 

At present, no tensor glueball candidate is promoted in this mass region  
but it was not always the case. Three out of four 
"firmly established" mesons listed in the tab.2 \cite{RPP(2010)}
\begin{equation} \label{54}
 f_2(2010),\quad f_2(2300),\quad f_2(2340) 
\end{equation} 
were discovered \cite{Etkin} in the eighties of the former century in 
the single experiment 
\begin{equation} \label{55}
 \pi^- p \longmapsto g_Tn\longmapsto \Phi \Phi n.  
\end{equation}
They were called $g_T$ mesons. Rich statistics of events justified the 
claim that the observed $\Phi\Phi$ signal descents from three separate 
mesons. 

The experiment raised big excitement since its authors claimed that the 
three mesons or "at least one of them" are glueballs. The belief was 
based on suggestion that $g_T$ mesons are created in the reaction which 
is doubly forbidden by OZI rule \footnote{The suggestion that the 
reaction (\ref{54}) is doubly OZI-forbidden induced  persistent 
opposition. When the dispute was prolonging excessively some 
people being interested in explaining the nature of the $g_T$ mesons  
but confused with this situation asked adversaries to work out a 
common conclusion at personal meeting. The appointment came face to 
face behind the closed doors. After this meeting an official statement 
was issued where the parties sustained their earlier positions \cite{LL}. 
So impasse in interpretation of the reaction (\ref{50}) was not overcome 
and interest to the nature of $g_T$ mesons gradually abated. However, 
the very existence of these mesons is not questioned and the masses 
measured in this experiment remain unchanged. Therefore, this event 
should not be forgotten. Possibly, now we shall have the opportunity to 
study structures of $g_T$ mesons without resolving the dispute on role 
of OZI rule in the reaction (\ref{55}).}.

Perhaps not all $g_T$ mesons are glueballs but if there 
exists any tensor glueball in this mass region then there should exist 
a decuplet as well. 
From two candidates  for the role of $x_2$ shown  in (\ref{51}) 
we choose this one which is "firmly established". This points out the 
meson $f_2(2010)$ as G-dominated state. Unfortunately, there is no 
justification in support of similar choice between the mesons (\ref{52}) 
and (\ref{53}). Therefore, it is impossible to fix particle content  
of the decuplet. However, it is good luck that we can indicate candidate 
to the role of G-dominated isoscalar meson thus supporting the existence 
of the decuplet 
\footnote{One can wonder whether all $g_T$ 
mesons may belong to the same decuplet \cite{KM}. If it were so then 
there should exist also further tensor mesons $a_2$ and $K_2$ satisfying 
MOR. So far such mesons were not observed}.

\section{Summary}
The existence of the glueball implies enlarging the meson nonet to the 
decuplet where the glueball state G is mixed with the isoscalar 
$q\bar{q}$ nonet states. The discovery of the G within the structure 
of decuplet helps in verification of its properties and makes the 
discovery more reliable. Therefore, decuplet occurs a right place for  
looking for it. The problem is in completing decuplet from the existing 
particles and describing it. We apply the VEC model. According to this 
model the multiplet is described by the master equations (ME). 
They form the specific sequence of equations depending only on the masses 
of the multiplet particles and the octet contents $l_i^2$ of the isoscalar 
physical states. The octet contents are the unknown variables of the 
sequence. Part of these equations (which is the basic system of the ME) 
determines the octet contents of the multiplet, the other ones  define 
its mass formulae (MF). The ME explain the genesis of the MF and relate 
their number to the number of ME. 

The number of ME is not given in advance as it is defined for each 
multiplet separately. To each number of ME there corresponds the 
individual multiplet of the physical particles. The multiplets having 
the same number of particles but different number of MF are considered 
as different. The existence of MF is not necessary for the multiplet 
existence.

The model well describes the nonets and predicts the existence of 
different kinds of them. The decuplets are defined just in the same way 
as other possible multiplets. The ME predict two kinds of decuplets. 
One of them has no MF, another has one MF. The latter is therefore more 
restrictive and its predictions are more definite. It puts many 
constraints on the masses. Especially useful is the mass ordering rule 
(MOR) which is transparent and very effective in selecting the particles 
to the decuplet. The MOR joins conditions of the particle existence with 
the requirement of satisfying the MF. 

The functions $l_i^2$, being solution of the basic sequence of ME, play 
important role in describing the multiplets.\\ 
1. They translate the natural positivity conditions $l_i^2>0$ (i=1,2,..) 
into the main restrictions on the masses of the nonets and decuplets. \\
2. They are the building blocks for constructing the mixing matrix as a 
function of masses showing at the same time room for ambiguities of 
its parameters.

The mixing matrix $V_G$ describes the decuplet of any domination pattern. 
It becomes completely determined if the "domination pattern" is indicated
\footnote{fixing the decuplet domination pattern plays the role similar 
to the (N,S) attributing the isoscalar states of the I-nonet}. 

The decuplet complying with the MF can be understood as a mixed state of 
the I-nonet and G. The admixture of G to the I-nonet transforms 
the pure (N, S) nonet solution of the ME into the decuplet functions 
$x_1$, $x_3$ which, together with $x_2$, constitute the solution of ME 
for decuplet. The masses of the $x_1$ and $x_3$ mesons of decuplet do not 
differ much from the masses of parental I-nonet (N, S). The mass gaps 
N - $x_1$ and $x_3$ - S are results of the G mixing. It is important that 
these gaps are smaller than the difference b - a in all known multiplets. 
That indicates that the coupling of G to $(q\bar{q})$ states is 
relatively weak. This, in turn, assures us that I-nonet (N, S) - pattern
of isoscalar mesons is preserved in the decuplet. 

The standard domination pattern of a decuplet (\ref{39}): 
\begin{eqnarray*}
x_1\sim N,    \quad    x_2\sim G,    \quad    x_3\sim S 
\end{eqnarray*}
is attributed to the decuplet of any $J^{PC}$ having one MF. The flavor 
wave functions of these decuplets are determined via the G-shaped mixing 
matrix (\ref{32}). 

This procedure is applied for the description of possible tensor 
decuplets. We present two candidates. 

One of the decuplets includes the mesons belonging to the currently 
known ground state nonet $2^{++}$ and the isoscalar meson $f_2(1430)$. 
The nonet isoscalar state $f_2(1270)$ is N-dominated and apparently its 
mass squared satisfies the relation $x_N<a$. Another isoscalar state 
$x_S$ which is S-dominated  has the mass squared close to b. For a long time 
only this nonet was considered as the multiplet of the $2^{++}$ mesons 
but now we find that the decuplet is also possible. This is due to the 
discovery of the anomalously large electromagnetic split of the 
$K^*_2(1430)$ mass. In this case the meson $f_2(1430)$ would serve 
as the lacking G-dominated state $x_2$. 
The candidate for the first tensor decuplet is
\begin{equation} \label{56} 
a_2(1320),\quad K_2^*(1430),\quad f_2(1270), \quad f_2(1430), 
\quad f_2'(1525).
\end{equation} 
The existence of the meson $f_2(1430)$ is uncertain. Its confirmation
would disclose the G-dominated structure of this meson as well as 
existence of the decuplet. However, such correlation between effects of the 
strong and electromagnetic phenomena would perhaps need some explanation.

Constructing another decuplet candidate is an exercise of selecting  
the mesons to the decuplet in the case of uncertain data. We propose   
\begin{equation} \label{57}
a_2(1700), K_2(1980), f_2(1565)/f_2(1640), f_2(1950)/f_2(2010), 
f_2(2300)/f_2(2340)
\end{equation}
which includes mesons listed in the tab.2. This decuplet enters the 
field of the long-standing controversy about nature of $g_T$ mesons. 

The G-dominated components of the tensor meson decuplets would be \\
$f_2(1430)$ and $f_2((1950)$/$f_2(2010)$.\\

We conclude with one more comment.\\
The VEC model describes meson multiplets of the broken $SU(3)$ flavor 
symmetry. The multiplets appear due to the difference between the masses 
of K and $\pi$ mesons (here we use pseudoscalar meson labels for any 
multiplet) and are treated in ME as an input. The difference between them 
is supposed to be an effect of the hard breaking of the $SU(3)$ symmetry. 
The mass splitting of the physical isoscalar decuplet mesons is regarded 
to be caused by other (soft) $SU(3)$ interaction between G, N, S states 
(the distinction of "very strong" and  "middle strong" interaction was 
recognized from very beginning of flavor  investigation). 
Soft interaction does not change the results of the hard 
one like big difference between the masses of K and $\pi$ mesons or 
determination of the domination pattern of the multiplet. 

Nonet and decuplet may have the same K and $\pi$ masses  
and identifying the right multiplet may appear difficult if 
not all isoscalar mesons are seen.   
Then we should  verify which of these multiplets better describes data.

\section{Conclusion}
The glueballs do exist. They can be found in the decuplets of mesons. 
If the decuplet obeys the mass formula the G-dominated state occupies 
central position between the isoscalar mesons $x_i$. The best 
candidates for glueball is $\eta (1405)$ and $f_2(1430)$ (if it exists).
Their advantage is that they belong to decuplets gathering the states 
with the most precisely measured masses.

\section{Acknowledgments}
Author thanks Professor Anna Urbaniak-Kucharczyk - Dean of the Physical 
Department of the University of Lodz; Professor  Pawe\l{} Ma\'slanka and 
Professor Jakub Rembieli\'nski - Chiefs of Theoretical Departments for 
their patient understanding and support; Professor  S.B. Gerasimov 
and Professor V.A. Meshcheryakov  (Dubna, Bogolyubov Laboratory of 
Theoretical Physics at Joint Institute of Nuclear Research) 
for durable friendly collaboration and any kind of help. Special thanks are 
expressed to Professor Piotr Kosi\'nski for many interesting discussions,
valuable comments and reading the manuscript; Dr Bartosz Zieli\'nski help 
in computer operations  is gratefully appreciated.

\section{Appendix 1. The model of vanishing exotic commutators (VEC)}
The following sequence of exotic commutators is assumed to vanish

\begin{equation}
    \left[T_a,\frac{d^{j}T_b}{dt^j}\right]=0,\quad \left(j=1,2,3,...\right)
\label{2.01}
\end{equation}
where $T$ is $SU(3)_F$ generator, $t$ is the time and $(a,b)$ is an
exotic combination of indices, i.e. such that the operator
$[T_a,T_b]$ does not belong to the octet representation.
Substituting $\frac{dT}{dt}=i[H,T]$, and using the infinite momentum
approximation for one-particle hamiltonian $H$ = $\sqrt{m^2+p^2}$
we transform equations (\ref{2.01}) into the system: \\
\begin{align}
    [T_a,[\hat{m^2},T_b]]&=0,\nonumber\\
    [T_a,[\hat{m^2},[\hat{m^2},T_b]]]&=0,\nonumber\\
    [T_a,[\hat{m^2},[\hat{m^2},[\hat{m^2},T_b]]]]&=0,\label{60}\\
    .....................................................&.......\nonumber
\end{align}
where $\hat{m^2}$ is the squared-mass operator.

For the matrix elements of the commutators (\ref{60}) between
one-particle states (we assume one-particle initial, final and
intermediate states) we obtain the sequence of equations involving
expressions $\langle x_8|(m^2)^j|x_8\rangle$ with different powers
$j=1,2,3,..$, where $x_8$ is the isoscalar state belonging to the
octet. Solving these equations, we obtain the sequence of formulae
for a multiplet of the light mesons. We have
\begin{equation} \label{61}
    \langle{x_8}\mid{\hat{(m^2)}^j}\mid{x_8}\rangle = \frac{1}{3} a^j +
    \frac{2}{3}b^j   \quad (j=1,2,3,...).    
\end{equation}
where $a$ is the mass squared of the isovector meson $a$; $b$ is
the mass squared of the subsidiary $s\bar{s}$ state,
\begin{equation}
    b=2K-a,    \label{2.04}
\end{equation}
and $K$, in turn, is the mass squared of the isospinor $K$ meson.

The isoscalar octet state $\mid{x_8}\rangle$ can be represented as
the linear combination of the physical isoscalar states
\begin{equation}
 \mid{x_8\rangle}=\sum{{l_i\mid{x_i}\rangle}}.
 \label{2.05}
\end{equation}
The coefficients $l_1$, $l_2$, $l_3$,.. determine octet contents of
the physical isoscalar states $|x_1\rangle$, $|x_2\rangle$,
$|x_3\rangle$,...  Substituting (\ref{2.05}) into (\ref{61}) we
obtain master equations (ME) of the multiplet:
\begin{equation}
    \sum{l_i^2x_i^j}=\frac{1}{3}a^j+\frac{2}{3}b^j, \quad (j=0,1,2,3,...)
    \label{2.06}
\end{equation}
where the $x_1$, $x_2$, $x_3$,... are isoscalar meson masses
squared. Normalization condition of the $l_i$ coefficients is
included into (\ref{2.06}) as equation for $j=0$.

\section{Appendix 2. Proof of MOR restrictions} 
The MOR restrictions for the decuplet masses have been formulated long 
ago in several approaches e.g. \cite{MT,enc} but they are most 
apparently seen from fig.1  of \cite{Where}. We give now their simple 
algebraic proof.

The functions $l_i^2$ being the solution of the ME (8) have to satisfy 
conditions of the octet content positivity (\ref{11}). 
Accepting the numbering of the isoscalar mesons (\ref{9}) we fix the 
signs of the denominators of the functions (\ref{20}). This determines 
the signs of their numerators:
\begin{subequations}\label{A}
\begin{align}
\label{A1}
(x_2-a)(x_3-a)+2(x_2-b)(x_3-b)>0,\\
\label{A2}
(x_1-a)(x_3-a)+2(x_1-b)(x_3-b)<0,\\
\label{A3}
(x_2-a)(x_1-a)+2(x_2-b)(x_1-b)>0.
\end{align}
\end{subequations}
By combining the inequalities (\ref{A}) with MF (\ref{21}) we can 
obtain several restrictions on the masses of the decuplet particles. 

In particular, assuming that $a>x_1$, i.e.
\begin{equation} \label{59}
(x_1-a)<0,
\end{equation}
multiplying (\ref{A1}) by (\ref{59}) and then subtracting equation 
(\ref{21})  we find, after omitting the positive factor (b - a), the 
inequality 
\begin{equation} \label{A60}
(x_2-b)(x_3-b)<0.
\end{equation}
Hence, according to (\ref{9})
\begin{equation} \label{A61}
(x_2-b)<0, \quad(x_3-b)>0. 
\end{equation} 
Repeating then the same operation with the product of (\ref{A3}) and 
(\ref{A61}) we find 
\begin{equation} \label{A62}
(x_1-a)(x_2-a)<0 
\end{equation}
and establish that 
\begin{equation} \label{A63}
(a-x_2)<0.
\end{equation}
The collection of the inequalities (\ref{59}), (\ref{A63}) and (\ref{A61})
form the rule (\ref{22}).\\
The rule holds  for any decuplet satisfying MF (\ref{21}).


\begin{thebibliography}{2}
\bibitem{Fri-Gel} H. Fritzsch, M. Gell-Mann, Proc.  the XVI Int.
Conf. on High Energy Physics, ({\bf12} Chicago-Batavia Ill.) 1972 vol. 2, 
p 135;
H. Fritzsch, P. Minkowski, {Nuovo Cim. A} {\bf30},393 (1975);
A. de Rujula, H. Georgi, S.L. Glashow {Phys. Rev. D}{\bf12} 147 (1975);
J. F. Willemsen, {Phys. Rev. D}{\bf13} 1327 (1976) 

\bibitem{Ros} J. L. Rosner, {Phys. Rev. D}{\bf24}, 1347 (1981)
\bibitem{Fr-N} P. G. O. Freund, Y. Nambu, {Phys. Rev. Lett.} {\bf34}, 
1645 (1975); 
N. Fuchs, {Phys. Rev. D} {\bf14}, 1912 (1976);
D. Robson, {Nucl. Phys} {\bf B130}, 1977 (1977)
\bibitem{R-T} J. L. Rosner, {Phys. Rev. D} {\bf27}, 1101 (1983); 
J. L. Rosner, S. F. Tuan, {Phys. Rev. D} {\bf27} 1544 (1983)
\bibitem{Mesh} S. Meshkov, {High Energy Physics}, Plenum Press. New York 
(1985) 
\bibitem{Se-Bi} F. Buisseret, V.  Mathieu, C. Semay {arXiiv: 0906.3098v2 [hep-ph]}; \\ 
P. Bicudo, S. Cotanch, F. J. Llanes-Estrada, D. G. Robertson 
{Eur. Phys.J. C}{\bf52} 363 (2007) [arXiv:hep-ph/06021]
  
\bibitem{Mor} C. J. Morningstar, M. Peardon, {Phys. Rev.D}{\bf60}, 034509 (1999);\\
 Y. Chen et al., {Phys. Rev. D}{\bf73}, 014516 (2006) 

\bibitem{MajT} M. Majewski, W. Tybor, {Acta Phys. Pol. B}{\bf15} 267 
(1984)
\bibitem{MT} M. Majewski, W. Tybor, {Acta Phys. Pol. B}{\bf15}, 777
(1984); Erratum, {Acta Phys. Pol. B}{\bf 15}, No 12 page 3 of the cover
\bibitem{Sum} M. Majewski, {Eur. Phys. J. C}{\bf30}, 223 (2003);
{hep-ph/0206285}
\bibitem{enc} M. Majewski, {Eur. Phys. J. C}{\bf46} 759 (2006); 
{hep-ph/0509008}
\bibitem{Where} M. Majewski, V. A. Meshcheryakov,  
{J. Phys. G  Nucl. Part. Phys} {\bf38} (2011) 035008; 
\bibitem{tensnonref}S. Godfrey and N. Isgur, {Phys. Rev. D}{\bf32} (1985) 189; 
L. Burakovsky and J. T. Goldman, Phys.Rev. D 57, 2879 (1998) 
[hep-ph/9703271]; 
D. -M. Li, H. Yu and Q. -X. Shen, J. Phys. G 27 (2001) 807 
[hep-ph/0010342]; 
C. -K. Chow and S. -J. Rey, JHEP9805 (1998) 010 [hep-ph/9708355]; 
F. Giacosa, T. Gutsche, V. E. Lyubovitskij, A. Faessler, Phys. Rev. D 72, 
114021 (2005) [hep-ph/0511171]; 
Z. -C. Ye,X. Wang, X. Liu and Q. Zhao, Phys. Rev. D 86 (2012) 054025 [
arXiv:1206.0097 [hep-ph]].
F. Buisseret, V. Mathieu, C. Semay arXiv: 0906.3098v2 [hep-ph] (2009)

\bibitem{RPP(2010)} Particle Data Group (Revew of Particle Properties 
(2014))  {Chinese Physics C} {\bf38} N9 (2014) 090001
\bibitem{AGST} C. Amsler, T. Gutsche, S. Spanier, N. A. Tornqvist,  
ibid p 630
\bibitem{AM} C. Amsler, G. L. Masoni. ibid  p 680
\bibitem{Etkin} A. Etkin et al., {Phys. Rev. Let.}{\bf40} 422 (1978);
{\bf49} 1620 (1982); 
A. Etkin et al.,{Phys. Lett B} {\bf165} 217 (1685); {Phys. Lett B} 
{\bf201} 568 (1988)

\bibitem{LL} H. J. Lipkin, {Phys. Lett. B} {\bf124} 509 (1983);
S. J. Lindenbaum, {Phys, Lett. B} {\bf131} 221 (1983); 
S. J. Lindenbaum, H.J. Lipkin, {Phys. Lett. B} {\bf149} 407 (1984)
\bibitem{KM} B. Kozlowicz, M. Majewski, {Acta Phys. Pol. B}{\bf20}
869 (1989)



\end{thebibliography}
\end{document}